\documentclass[pra,preprint,eqsecnum,showpacs]{revtex4}
\usepackage{amsmath}
\usepackage{amsfonts}
\usepackage{amssymb}
\usepackage{graphicx}
\usepackage{bm}
\begin{document}
\title{The relativistic massless harmonic oscillator}
\author{K. Kowalski and J. Rembieli\'nski}
\affiliation{Department of Theoretical Physics, University
of \L\'od\'z, ul.\ Pomorska 149/153, 90-236 \L\'od\'z,
Poland}
\begin{abstract}
A detailed study of the relativistic classical and quantum mechanics of
the massless harmonic oscillator is presented.
\end{abstract}
\maketitle
\section{Introduction}
One of the most important physical systems for both classical
and quantum mechanics is the harmonic oscillator.  In
contrast to the nonrelativistic harmonic oscillator that is discussed in
most textbooks, the theory of relativistic harmonic oscillator is far
from complete.  The reason for this is the complexity of the problem related to
the nonlinearity of differential equations of motion for the
classical relativistic oscillator.  No wonder that even in the
simple case of the massive one-dimensional relativistic oscillator
there are problems with identification of periodic solutions to
equations of motion \cite{1}.  The problem of a quantum relativistic harmonic 
oscillator is usually formulated in one of three different  frameworks: 
the Klein-Gordon, Dirac or Salpeter equations.  The first one uses the spinless 
Klein-Gordon equation with a Lorentz invariant oscillatory potential \cite{2}. 
However, the solutions of that equation are blamed 
by pathologies such as the appearance of ghost states. The second 
approach, referred to by Moshinsky \cite{3} as the ``Dirac oscillator" and describing 
spin one-half particles utilizes the Dirac equation with an appropriate combination 
of the scalar, vector and tensor couplings with an external field \cite{4}. It can 
be successively applied to analysis of relativistic symmetries which recently were 
recognized experimentally in both nuclear and hadron spectroscopy \cite{5}.   
Unfortunately, this approach has no classical relativistic counterpart.  Finally, 
the third approach follows from the relativistic Hamiltonian dynamics for a scalar 
particle and on the quantum level it is based on the spinless Salpeter equation \cite{6}. 
The Salpeter equation \cite{6,7,8,9,10,11,12,13,14,15,16} is a "square root" of the 
Klein-Gordon equation \cite{17} and can be regarded as its alternative \cite{16}.  
The serious advantages of the Salpeter scheme are the lack of problems with probabilistic 
interpretation on the quantum level as well as the classically well-defined physical 
content of this theory. This last framework is frequently used as  a phenomenological 
description of the quark-antiquark-gluon system as a hadron model. 

Surprisingly, to our best knowledge,
the simplest case of the massless relativistic harmonic oscillator
was not discussed in the literature.  In this work we perform a 
detailed analysis of the massless relativistic harmonic oscillator.
In particular we find the exact solutions to the classical Hamilton 
equations as well as to the corresponding quantum Salpeter equation 
and discuss their basic properties.  The article is organized as follows.  
In Sec.\ II, by integrating the corresponding Hamilton system we identify all
kinds of possible motion of the oscillator as well as find its
quantative characteristics.  For an easy illustration of the
dynamics of the relativistic massless harmonic oscillator we also
provide a graphical presentation of numerical integration of
equations of motion.  Section III is devoted to the
quantization of the massless relativistic harmonic oscillator.  
\section{The analysis of the classical relativistic massless
harmonic oscillator}
The Hamiltonian of the relativistic massless particle subject to the
potential $\frac{1}{2}\kappa^2\bm{x}^2$ is given by
\begin{equation}
H = c|{\bm p}| + \frac{1}{2}\kappa^2{\bm x}^2,
\end{equation}
where ${\bm x}$ and ${\bm p}$ are the position and the momentum of a
particle, $|{\bm p}|=\sqrt{{\bm p}^2}$ is the norm of the vector
${\bm p}$ (so $c|{\bm p}|$ is the kinetic energy of the particle), 
$\kappa$ is a constant and $c$ is the speed of light.  
Therefore, the Hamilton's equations are
\begin{eqnarray}
\dot {\bm x} &=& c\frac{{\bm p}}{|{\bm p}|},\nonumber\\
\dot {\bm p} &=& -\kappa^2{\bm x}.\nonumber\\
\end{eqnarray}
We point out that an immediate consequence of Eq.\ (2.2) is ${\dot{\bm
x}}^2=c^2$, that is the length of velocity is $c$ as should be for a
massless particle.  The familiar integrals of the motion in a
central field \cite{18} are the energy $E$ and the angular momentum ${\bm J}$:
\begin{eqnarray}
E &=& c|{\bm p}|+\frac{\kappa^2}{2}{\bm x}^2,\\
{\bm J} &=& {\bm x}\times{\bm p}.
\end{eqnarray}
As a result of the conservation of the angular momentum ${\bm J}$ the
motion is planar and we can restrict, without loss of 
generality, to the case of a particle moving in the $(x^1,x^2)$ plane.
On passing to the polar coordinates ${\bm x}=(x^1,x^2)=(r\cos\varphi,
r\sin\varphi)$ and ${\bm p}=(p^1,p^2)=(p\cos\theta,p\sin\theta)$, where
$r=|{\bm x}|$, and $p=|{\bm p}|$, we obtain from Eq.\ (2.2) the following
system:
\begin{eqnarray}
\dot r &=& c\cos(\theta-\varphi),\nonumber\\
\dot\varphi &=& \frac{c}{r}\sin(\theta-\varphi),\nonumber\\
\dot p &=& -\kappa^2r\cos(\theta-\varphi),\nonumber\\
\dot\theta &=& \kappa^2\frac{r}{p}\sin(\theta-\varphi).\nonumber\\
\end{eqnarray}
The integrals of the motion take the form
\begin{eqnarray}
E &=& cp+\frac{1}{2}\kappa^2r^2,\\
J &\equiv& J_3 = rp\sin(\theta-\varphi).
\end{eqnarray}
From (2.5), (2.6) and (2.7) we find
\begin{equation}
r\sqrt{r^2\left(E-\frac{\kappa^2}{2}r^2\right)^2-(Jc)^2}\,\,d\varphi=\pm |J|c dr.
\end{equation}
We point out that the ``$+$'' and ``$-$'' signs correspond to the
two possible orientations of the angular momentum. We choose, without
loss of generality, the sign ``$+$'' and $J>0$ throughout this
work.  Now, from Eq.\ (2.8) we find that the trajectories should satisfy
\begin{equation}
r\left(E-\frac{\kappa^2}{2}r^2\right)\ge Jc,
\end{equation}
and we can classify the types of motion as folows.  The first
two possibilities refer to $J\ne 0$.  Namely,\bigskip\\
\noindent 1)\quad For $r\left(E-\frac{\kappa^2}{2}r^2\right) = Jc$, we get
\begin{equation}
r=R=\sqrt{\frac{2E}{3\kappa^2}},\qquad \varphi=\omega
t+\varphi_0,\qquad p=p_0=\frac{2E}{3c},\qquad \theta =\omega
t+\varphi_0+\frac{\pi}{2},
\end{equation}
where $\omega
=\frac{c}{R}=\frac{\kappa^2R}{p_0}=c\sqrt{\frac{3\kappa^2}{2E}}$.
So in this case we have a uniform motion in a circle with the linear
speed $|{\bm v}|=\omega R=c$.  This solution can also be obtained from
Eq.\ (2.2) by demanding ${\bm x}\mbox{\boldmath${\cdot}$}{\bm p}=0$.  
Indeed, it can be easily checked that (2.2) and (2.3) imply the system
\begin{eqnarray}
&&\frac{d}{dt}{\bm x}\mbox{\boldmath${\cdot}$}{\bm p} =
E-\frac{3}{2}\kappa^2{\bm x}^2,\nonumber\\
&&\frac{d}{dt}{\bm p}^2 = -2\kappa^2{\bm
x}\mbox{\boldmath${\cdot}$}{\bm p},\nonumber\\
&&\frac{d}{dt}{\bm x}^2 = \frac{2c}{|{\bm p}|}{\bm
x}\mbox{\boldmath${\cdot}$}{\bm p}.\nonumber\\
\end{eqnarray}
From (2.11) it follows easily that the orthogonality of ${\bm x}$
and ${\bm p}$ refers to the motion of a particle in a circle with radius
$|{\bm x}|=R=\sqrt{\frac{2E}{3\kappa^2}}$.\bigskip\\
\noindent 2)\quad For $r\left(E-\frac{\kappa^2}{2}r^2\right)> Jc$,
the path lies entirely within the annulus bounded by the circles
$r=r_{\rm min}$ and $r=r_{\rm max}$, that is we have
\begin{equation}
r_{\rm min}\le r\le r_{\rm max},
\end{equation}
where $r_{\rm min}$ and $r_{\rm max}$ are the real positive
solutions of the equation
\begin{equation}
\frac{\kappa^2}{2}r^3 -rE+Jc=0.
\end{equation}
We find after some calculation
\begin{eqnarray}
r_{\rm min} &=& 2\sqrt{\frac{2E}{3\kappa^2}}\sin\frac{\alpha}{3},\\
r_{\rm max} &=&
\sqrt{\frac{2E}{3\kappa^2}}(\sqrt{3}\cos\frac{\alpha}{3}-\sin\frac{\alpha
}{3}),
\end{eqnarray}
where $\sin\alpha
=\frac{Jc}{\kappa^2}\left(\frac{3\kappa^2}{2E}\right)^\frac{3}{2}$,
and $0\le\alpha \le\frac{\pi}{2}$.

We now return to Eq.\ (2.8).  An immediate consequence of integration of 
Eq.\ (2.8) is the the relation
\begin{equation}
\varphi = \varphi_0 +\frac{Jc}{2}\int_{r_0^2}^{r^2}\frac{dx}
{x\sqrt{x(E-\frac{\kappa^2}{2}x)^2-(Jc)^2}}.
\end{equation}
In the particular case of $r_0=r_{\rm min}\ne 0$, and $r(t)>r_{\rm
min}$, $t>0$, the integral from the right-hand side of (2.16) can be
expressed by means of the elliptic integral of the third kind
$\Pi(\phi,n,k)$ (see Ref.\ \cite{19}, 3.137, Eq.\ 3), namely we have
\begin{equation}
\varphi=\varphi_0+\frac{2Jc}{\kappa^2r_{\rm min}^2\sqrt{r_-^2-r_{\rm min}^2}}
\Pi\left(\arcsin\sqrt{\frac{r^2-r_{\rm min}^2}{r_{\rm max}^2-r_{\rm
min}^2}},1-\frac{r_{\rm max}^2}{r_{\rm min}^2},\sqrt{\frac{r_{\rm max}^2
-r_{\rm min}^2}{r_-^2-r_{\rm min}^2}}\right),\quad r_0=r_{\rm min}.
\end{equation}
where $r_-$ is the negative root of the polynomial from the left-hand
side of Eq.\ (2.13) satisfying
\begin{equation}
r_{\rm min}+r_{\rm max}+r_-=0,\qquad r_-^2>r_{\rm max}^2>r_{\rm min}^2.
\end{equation}
Clearly, Eq.\ (2.17) defines $r$ as an implicit function of $\varphi$.

Furthermore, for $r_0=r_{\rm max}\ne 0$, and $r(t)<r_{\rm max}$,
$t>0$, the implicit equation for the trajectory can be obtained from
(2.16) with the help of the elliptic functions of the third kind
$\Pi(\phi,n,k)$ and first kind $F(\phi,k)$ (see \cite{19}, 3.137, Eq.\
4).  It follows that
\begin{eqnarray}
&&\varphi = \varphi_0-\frac{2Jc}{\kappa^2r_-^2r_{\rm max}^2\sqrt{r_-^2-r_{\rm min}^2}}
\Bigg[(r_-^2-r_{\rm max}^2)\nonumber\\
&&{}\times\Pi\left(
\arcsin\sqrt{\frac{(r_-^2-r_{\rm min}^2)(r_{\rm max}^2-r^2)}{(r_{\rm
max}^2-r_{\rm min}^2)(r_-^2-r^2)}},
\frac{r_{\rm max}^2-r_{\rm min}^2}{r_-^2-r_{\rm min}^2}
\frac{r_-^2}{r_{\rm max}^2},
\sqrt{\frac{r_{\rm max}^2-r_{\rm min}^2}{r_-^2-r_{\rm min}^2}}\right)\nonumber\\
&&{}+r_{\rm max}^2F\left(\arcsin\sqrt{\frac{(r_-^2-r_{\rm min}^2)(r_{\rm max}^2-r^2)}
{(r_{\rm max}^2-r_{\rm min}^2)(r_-^2-r^2)}},\sqrt{\frac{r_{\rm max}^2-r_{\rm min}^2}
{r_-^2-r_{\rm min}^2}}\right)\Bigg],\quad r_0=r_{\rm max}.\nonumber\\
\end{eqnarray}
Now, the four-dimensional system (2.2), where ${\bm x}=(x^1,x^2)$,
and ${\bm p}=(p^1,p^2)$ is completely integrable.  Indeed, it
possesses two integrals in involution $E$ and $J$.  Therefore, the
motion between two circles with the radius $r_{\rm min}$ and $r_{\rm
max}$ can be only quasiperiodic and periodic.  Of course the case
of the periodic motion refers to a closed path.  This means that an
angle $\Delta\varphi$ given by (see formula (2.17))
\begin{eqnarray}
\Delta\varphi &=& \frac{Jc}{2}\int_{r_{\rm min}^2}^{r_{\rm max}^2}\frac{dx}
{x\sqrt{x(E-\frac{\kappa^2}{2}x)^2-(Jc)^2}}\nonumber\\
{} &=& \frac{2Jc}{\kappa^2r_{\rm min}^2\sqrt{r_-^2-r_{\rm min}^2}}
\Pi\left(\frac{\pi}{2},1-\frac{r_{\rm max}^2}{r_{\rm min}^2},\sqrt{\frac{r_{\rm max}^2
-r_{\rm min}^2}{r_-^2-r_{\rm min}^2}}\right),
\end{eqnarray}
should be a rational function of $\pi$, i.e.\ $\Delta\varphi=\pi
m/n$, where $m$ and $n$ are integers.  An example of a periodic
motion between two circles is presented in Figs 1 and 2.  It
should be noted that the length of momentum (kinetic energy of the
particle $c|{\bm p}|$) has maximum at $r=r_{\rm
min}$, decreases (increases) as $r$ approaches $r_{\rm max}$ ($r_{\rm min}$), 
and for $r=r_{\rm max}$ has minimum.  Clearly, such behavior of the momentum of a
massless particle is consistent with the form of (2.6).  The values
of $r_{\rm min}$ and $r_{\rm max}$ as well as extrema of the length
of momentum can be expressed as a function of the energy by means of
the implicit formulas (2.14) and (2.15).  We finally
remark that the case of the uniform motion in a circle discussed
earlier [type 1) of the motion] refers to the condition $r_{\rm
min}=r_{\rm max}=R=\sqrt{\frac{2E}{3\kappa ^2}}$.
\begin{figure*}
\centering
\includegraphics[scale=.8]{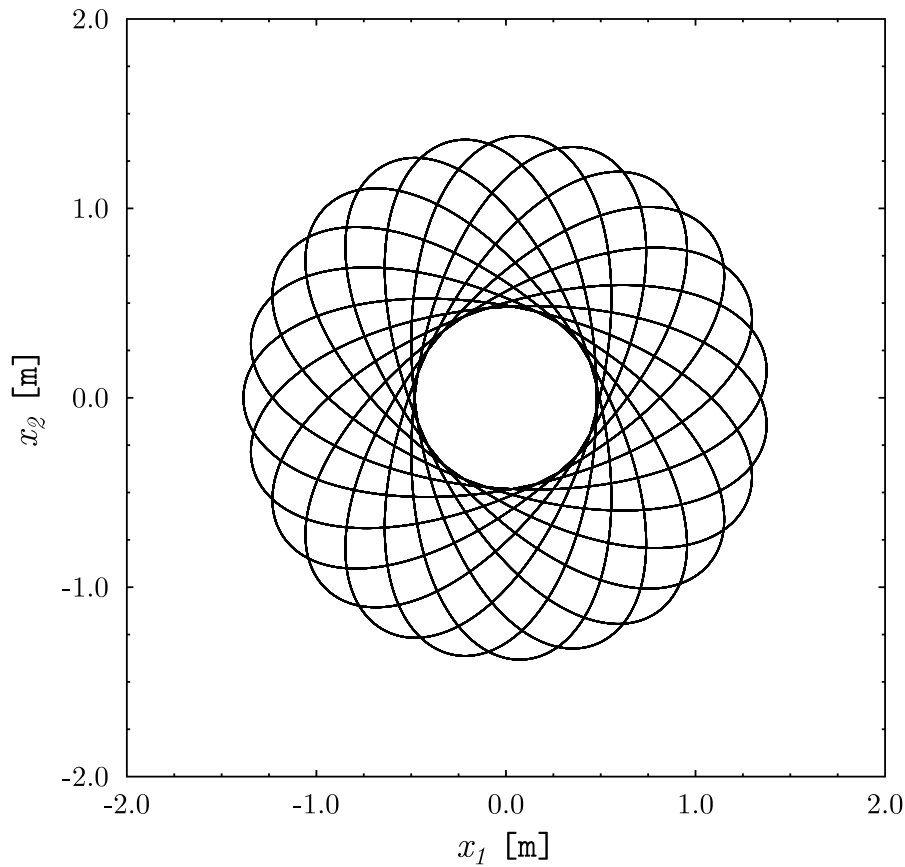}
\caption{The periodic solution of the system (2.2) obtained
by numerical integration.  The initial data are ${\bm x}_0=(0.479000,0.000000)$ m, 
${\bm p}_0=(0.000000,1.290805)$ Js${\rm m}^{-1}$  , the parameter $\kappa^2=1$ 
J${\rm m}^{-2}$, and $c=1$ m${\rm s}^{-1}$.}
\end{figure*}
\begin{figure*}
\centering
\includegraphics[scale=.8]{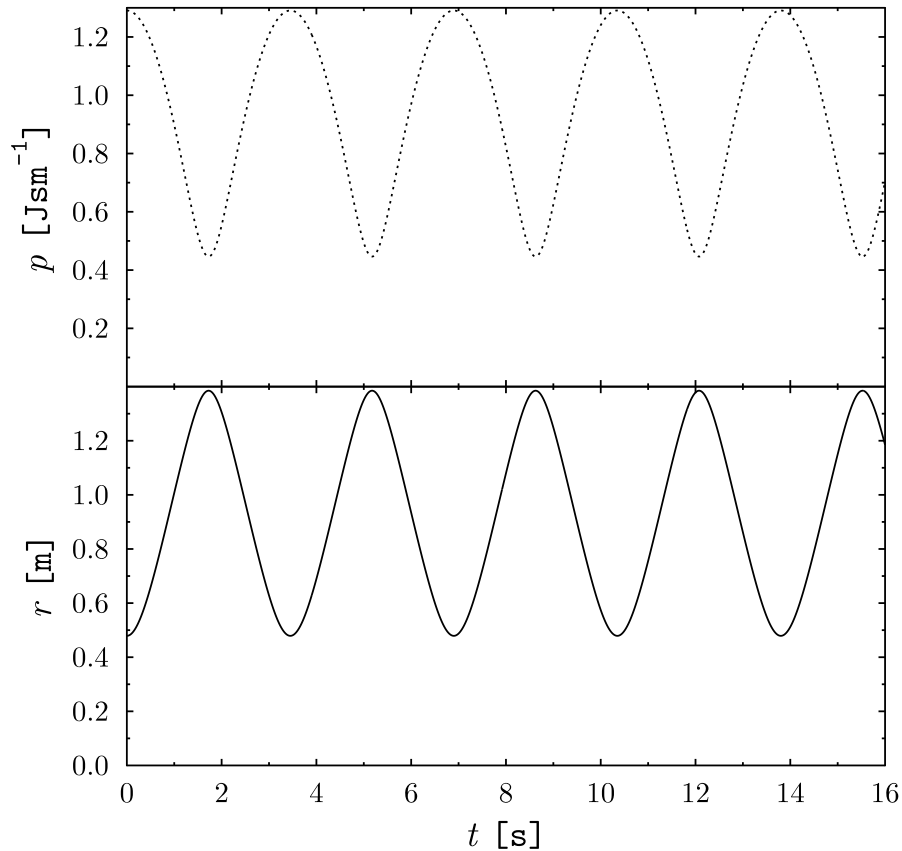}
\caption{The plot of the radius $r=|{\bm x}|$ (solid line) and the length of momentum
$p=|{\bm p}|$ (dotted line) vs time obtained by numerical integration of (2.11).  The initial 
condition is the same as in Fig.\ 1.}
\end{figure*}

The third type of the motion corresponds to $J=0$, so we
have\bigskip\\
\noindent 3)\quad $r(E-\frac{\kappa^2}{2}r^2)\ge 0$.  From this
inequality we find $0\le r\le r_{\rm max}=\frac{\sqrt{2E}}{\kappa}$.
On the other hand, taking into account (2.7) we find that for $J=0$
the system (2.5) reduces to
\begin{eqnarray}
\dot r &=& \pm c,\nonumber\\
\dot\varphi &=& 0,\nonumber\\
\dot p &=& \mp\kappa^2r,\nonumber\\
\dot\theta &=& 0,\nonumber\\
\end{eqnarray}
where $\theta-\varphi=0$ or $|\theta-\varphi|=\pi$.  Therefore a
particle motion is uniform in a segment
$[0,\frac{\sqrt{2E}}{\kappa}]$, more precisely, we have $r=\pm ct+r_0$,
where $0\le r\le \frac{\sqrt{2E}}{\kappa}$ and the two signs correspond
to two possible directions of motion.  Assuming that a particle
moves in the $x$-coordinate line, (i.e.\ $x=x^1$), we get
\begin{equation}
x=\pm ct+x_0,\qquad -r_{\rm max}\le x\le r_{\rm max},
\end{equation}
where the turning points are $x=r_{\rm max}$ and $x=-r_{\rm
max}$.  On setting $x_0=-r_{\rm max}$ we can write the trajectory
explicitly as
\begin{equation}
x(t) =
(-1)^{\left[\frac{2t}{T}\right]}c\left\{t-\left(2\left[\frac{2t}{T}\right]+
1\right)\frac{T}{4}\right\},
\end{equation}
where $T=\frac{4r_{\rm max}}{c}$ is the period of oscillations of a
massless particle between the turning points $x=r_{\rm max}$ and 
$x=-r_{\rm max}$, and $[a]$ is the biggest integer in $a$.  The
trajectory (2.23) is illustrated in Fig.\ 3.
Notice that at the turning points the momentum of a massless particle
vanishes [see Eq.\ (2.6) for $r=r_{\rm max}$] that is $p_{\rm min}=0$.  The 
maximum value of momentum $p_{\rm max}=\frac{E}{c}$ is reached for $x=0$.
The time development of the momentum for $p_0=0$ and $x_0=-r_{\rm max}$ 
can be written in the form
\begin{equation}
p=-\frac{\kappa^2}{2}\left(x^2(t)-c\frac{T}{4}\right),
\end{equation}
where $x(t)$ is given by (2.23).  The plot of $p$ versus $t$ is shown in 
Fig.\ 3.  Because the momentum
of a massless particle tends to zero as its position approaches the turning
point we deal with a ``red shift'' similar to the gravitational
one.  
\begin{figure*}
\centering
\includegraphics[scale=.8]{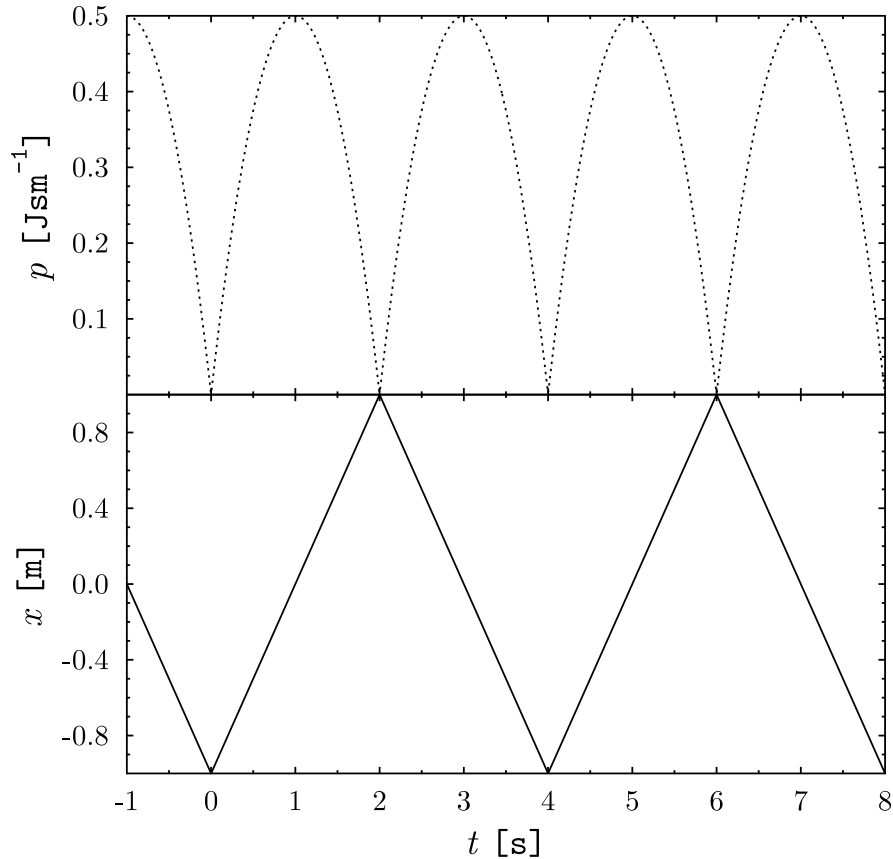}
\caption{The plot of the coordinate (solid line) and the momentum (dotted line)
of a massless oscillating particle vs time given by (2.23) and (2.24), respectively, 
where $c=1$ m${\rm s}^{-1}$, $\kappa^2=1$ J${\rm m}^{-2}$, $r_{\rm max}=1$ m, 
and $T=4$ s.}
\end{figure*}
It should also be noted that the motion in the segment can be 
easily obtained from (2.11) by setting $\frac{{\bm x}\mbox{\boldmath$
\scriptstyle{\cdot}$}{\bm p}}{|{\bm x}||{\bm p}|}=\pm 1$, that is 
${\bm x}$ and ${\bm p}$ are parallel or antiparallel and therefore 
satisfy ${\bm x}\times{\bm p} ={\bm 0}$.  
Evidently, in the case of the system (2.2) this 
condition is equivalent to $J=0$.  We point out that the motion 
in a segment corresponds to the condition $r_{\rm min}=0$ 
and $r_{\rm max}=\frac{\sqrt{2E}}{\kappa}$ for the nonnegative 
solutions to (2.13).  We finally remark that the type of motion
is completely determined by the values of the energy $E$ and the
angular momentum $J$.  Namely, using the parametrization of $r_{\rm
min}$ and $r_{\rm max}$ defined by (2.14) and (2.15) we find
\begin{equation}
0\le \frac{Jc}{\kappa^2}\left(\frac{3\kappa^2}{2E}\right)^{\frac{3}{2}}
\le 1,
\end{equation}
where
$\frac{Jc}{\kappa^2}\left(\frac{3\kappa^2}{2E}\right)^{\frac{3}{2}}=1$
refers to the motion in a circle,
$0<\frac{Jc}{\kappa^2}\left(\frac{3\kappa^2}{2E}\right)^{\frac{3}{2}}<1$
corresponds to the motion between two circles, and
$\frac{Jc}{\kappa^2}\left(\frac{3\kappa^2}{2E}\right)^{\frac{3}{2}}=0$,
i.e.\ $J=0$ is the condition for the motion in the segment.
\section{Quantum mechanics of the relativistic massless harmonic
oscillator}
In relativistic quantum mechanics the massless harmonic oscillator 
defined by the Hamiltonian (2.1) is described by a massless version of 
the spinless Salpeter equation
\begin{equation}
{\rm i}\hbar\frac{\partial}{\partial t}\psi({\bm x},t)=
\left(c\hbar\sqrt{-\Delta_{\bm x}}+\frac{\kappa^2}{2}{\bm
x}^2\right)\psi ({\bm x},t),
\end{equation}
where $\Delta_{\bm x}=(\frac{\partial}{\partial{\bm x}})^2$.
Therefore the eigenvalue equation for the Hamiltonian 
$\hat H\psi_E =E\psi_E$ takes the form of the pseudodifferential 
equation
\begin{equation}
\left(c\hbar\sqrt{-\Delta_{\bm x}}+\frac{\kappa^2}{2}{\bm
x}^2\right)\psi_E({\bm x})=E\psi_E({\bm x}).
\end{equation}
Performing the Fourier transformation
\begin{equation}
\psi({\bm x)}=\frac{1}{(2\pi\hbar)^\frac{3}{2}}\int d^3{\bm
k}e^{{\rm i}\frac{{\bm k}\mbox{\boldmath$\scriptstyle{\cdot}$}
{\bm x}}{\hbar}}\tilde\psi({\bm k}),
\end{equation}
we get from (3.2) the following equation:
\begin{equation}
\left(-\Delta_{\bm k}+\frac{2c}{(\kappa\hbar)^2}|{\bm
k}|\right)\tilde\psi_E({\bm k})=\frac{2E}{(\kappa\hbar)^2}
\tilde\psi_E({\bm k}), 
\end{equation}
where $\Delta_{\bm k}=(\frac{\partial}{\partial{\bm k}})^2$.
Finally, switching over to the spherical coordinates ${\bm
k}=(k\sin\alpha\cos\beta,k\sin\alpha\sin\beta,k\cos\alpha)$, where
$k=|{\bm k}|$, and making the ansatz
\begin{equation}
\tilde\psi_E({\bm k})=\frac{\chi(k)}{k}Y^m_l(\alpha,\beta),
\end{equation}
where $Y^m_l(\alpha,\beta)$ are the spherical functions, we obtain
the ``radial equation''
\begin{equation}
\left(-\frac{d^2}{dk^2}+\frac{l(l+1)}{k^2}+\frac{2c}{(\kappa\hbar)^2}k
\right)\chi(k)=\frac{2E}{(\kappa\hbar)^2}\chi(k).
\end{equation}
To our best knowledge in the case of $l\ne0$ the solution of (3.6) is 
not known.  For $l=0$ the solution to (3.6) can be expressed by
means of the Airy function ${\rm Ai}(x)$ \cite{20}, namely
\begin{equation}
\chi(k) =
C{\rm Ai}\left[\frac{2c}{(2c\kappa\hbar)^\frac{2}{3}}\left(k-\frac{E}{c}\right)\right],
\end{equation}
where $C$ is constant.  We point out that $l=0$ was also the case
discussed in \cite{6}, where the recurrence was identified
satisfied by coefficients of the formal power series expansion for the
solution to the spinless Salpeter equation corresponding to the massive
relativistic harmonic oscillator.  Clearly, $l=0$ refers to the vanishing
angular momentum, therefore we deal in this case with the
quantization of the motion of a massless particle in the segment
$0\le r\le r_{\rm max}=\frac{\sqrt{2E}}{\kappa}$ discussed in the 
previous section corresponding to the condition $J=0$ (third type of
the motion).  Furthermore, for $l=0$ the ansatz (3.5) takes the form
\begin{equation}
\tilde\psi_E({\bm k})=
\frac{\chi(k)}{k}Y^0_0(\alpha,\beta)=\frac{1}{\sqrt{4\pi}}\frac{\chi(k)}{k}.
\end{equation}
Demanding that $\tilde\psi_E({\bm k})$ is well defined for $k=0$
we find $\chi(0)=0$ (compare \cite{21} Eq.\ (32.11)), which leads to
${\rm Ai}\left(-\frac{2E}{(2c\kappa\hbar)^\frac{2}{3}}\right)=0$.  This
quantization condition means that the values of the energy $E_n$,
$n=1,2,\ldots$, are given by zeros of the {\rm Ai}ry function $a_n$.  We
have
\begin{equation}
E_n = -\frac{(2c\kappa\hbar)^\frac{2}{3}}{2}a_n,\qquad n=1,2,\ldots .
\end{equation}
Using the fact that the functions ${\rm Ai}(x+a_n)/{\rm Ai}'(a_n)$,
$n=1,2,\ldots$, where ${\rm Ai}'(x)$ designates the derivative of the
Airy function ${\rm Ai}(x)$, form an orthonormal basis on the interval
$[0,\infty)$ \cite{16}, we find that the normalized solutions (3.5)
to (3.4) in the Hilbert space $L^2({\Bbb R}^3,d^3{\bm k})$, with $l=0$ 
can be written as
\begin{equation}
\tilde\psi_n({\bm k})\equiv \tilde\psi_{E_n}({\bm k}) =
\sqrt{\frac{c}{2\pi}}\frac{1}{(2c\kappa\hbar)^\frac{1}{3}}\frac{1}
{{\rm Ai}'(a_n)}\frac{1}{k}{\rm
Ai}\left(\frac{2c}{(2c\kappa\hbar)^\frac{2}{3}}k+a_n\right).
\end{equation}
From (3.10) and (3.3) we finally obtain the normalized wave functions
such that
\begin{equation}
\psi_n({\bm x}) =
\sqrt{\frac{c}{\hbar}}\,\frac{1}{\pi}\frac{1}{(2c\kappa\hbar)^\frac{1}{3}}\frac{1}
{{\rm Ai}'(a_n)}\frac{1}{r}\int_{0}^{\infty}dk\sin\frac{kr}{\hbar}{\rm
Ai}\left(\frac{2c}{(2c\kappa\hbar)^\frac{2}{3}}k+a_n\right),
\end{equation}
where $r=|{\bm x}|$.  

As in the case of the nonrelativistic harmonic
oscillator with the probability density different from zero outside
the turning points, the probability density $\rho_n(r)=|\psi_n({\bm
x})|^2$ does not vanish for $r>r_{\rm max}(E_n)$, where $r_{\rm
max}(E_n)=\frac{\sqrt{2E_n}}{\kappa}$, $n=1,2,\ldots,$.  However, it
follows from the numerical calculation that $\rho_n(r)$ has no
maxima for $r>r_{\rm max}(E_n)$ (see Fig.\ 4).  Furthermore, taking 
into account all directions of the motion in the segment $[0,r_{\rm max}]$
(classical limit does not deal with a single classical orbit but an
ansamble of classical orbits \cite{22}) and taking into account that the
probability of finding a particle in the spherical layer $r$, $r+dr$
is inverse proportional to the surface of the sphere with radius
$r$, we find that the classical probability density is given by the
formula
\begin{equation}
\rho_{\rm cl}({\bm x}) \equiv \rho_{\rm cl}(r)= \frac{\theta(r_{\rm
max}-r)}{4\pi r_{\rm max}r^2},
\end{equation}
where $r_{\rm max}=\frac{\sqrt{2E}}{\kappa}$ and $\theta(x)$ is the
Heaviside step function.  Clearly, the normalization 
condition is of the form
\begin{equation}
\int d^3{\bm x}\rho_{\rm cl}({\bm x})=\int_0^\infty
\rho_{\rm cl}(r)d\mu(r) = 1,
\end{equation}
where $d\mu(r)=4\pi r^2dr$.  The comparison of the quantum
probability density $\rho_n(r)$, and the classical one $\rho_{\rm
cl}(r)$ for $r_{\rm max}(E_n)$ is shown in Fig.\ 4.  As expected the 
differences between the quantum and the classical descriptions decrease 
as the quantum number $n$ increases.
\begin{figure*}
\centering
\begin{tabular}{c@{}c}
\includegraphics[width =.5\textwidth]{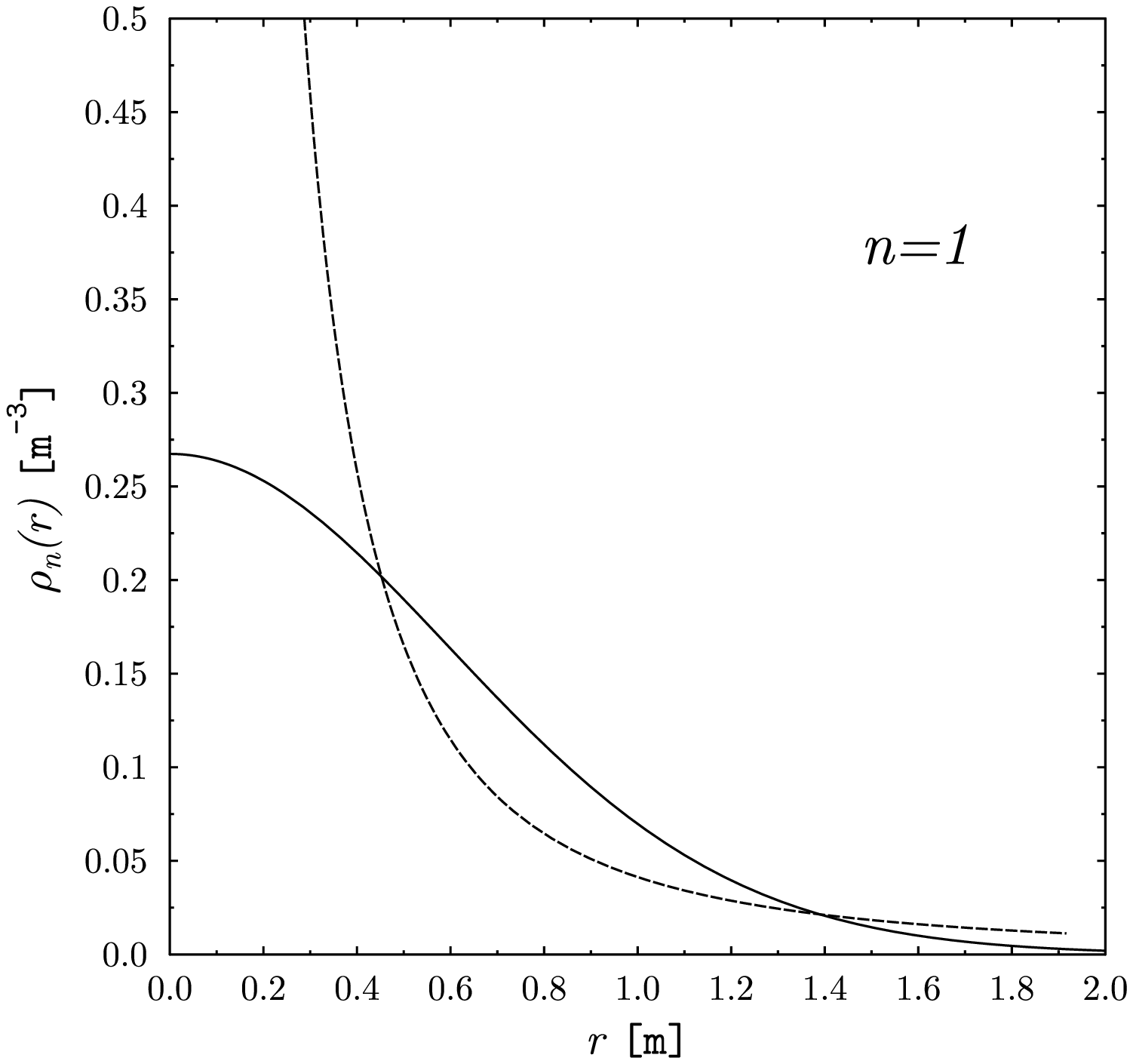}&
\includegraphics[width =.5\textwidth]{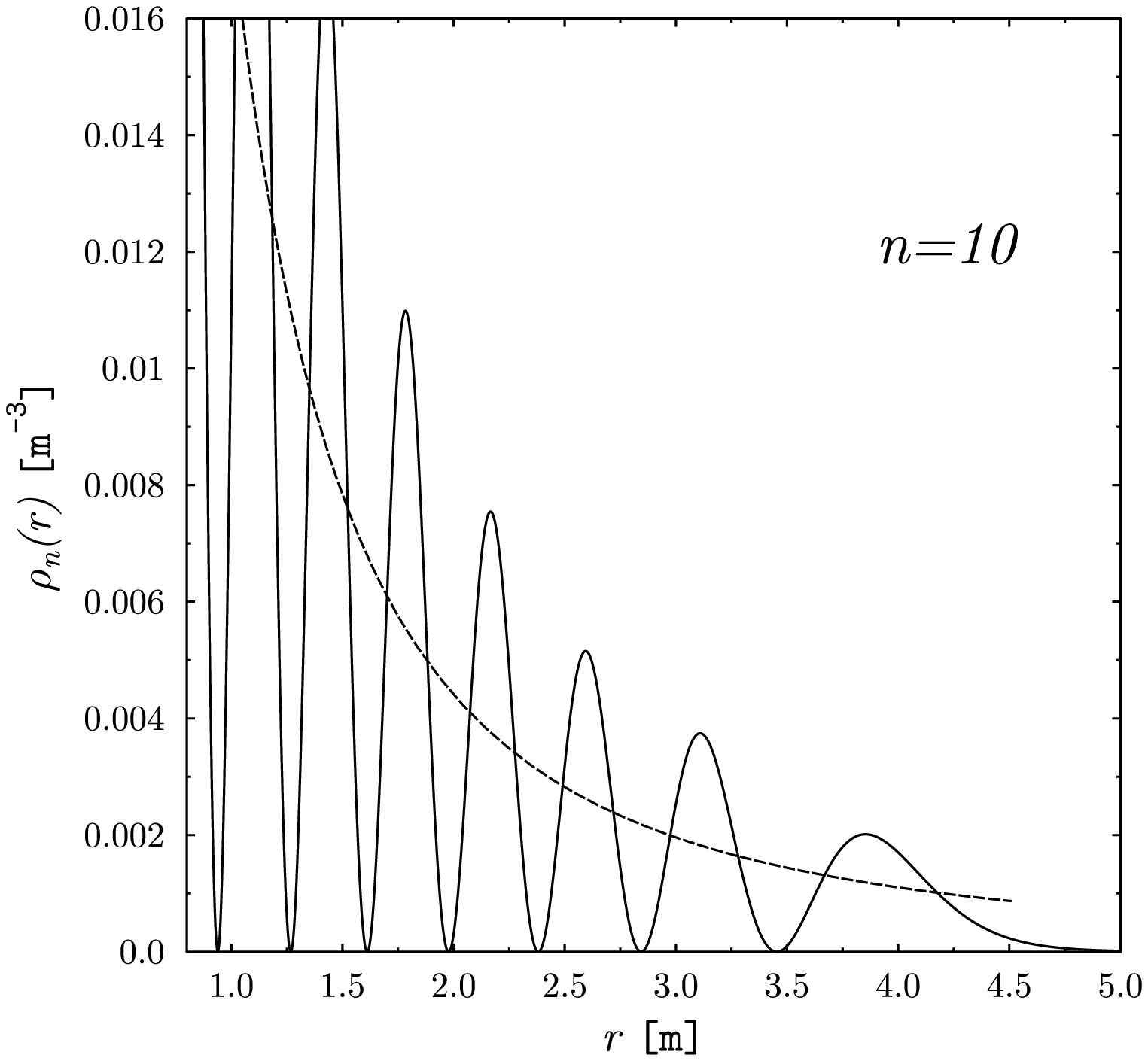}\\
\includegraphics[width =.5\textwidth]{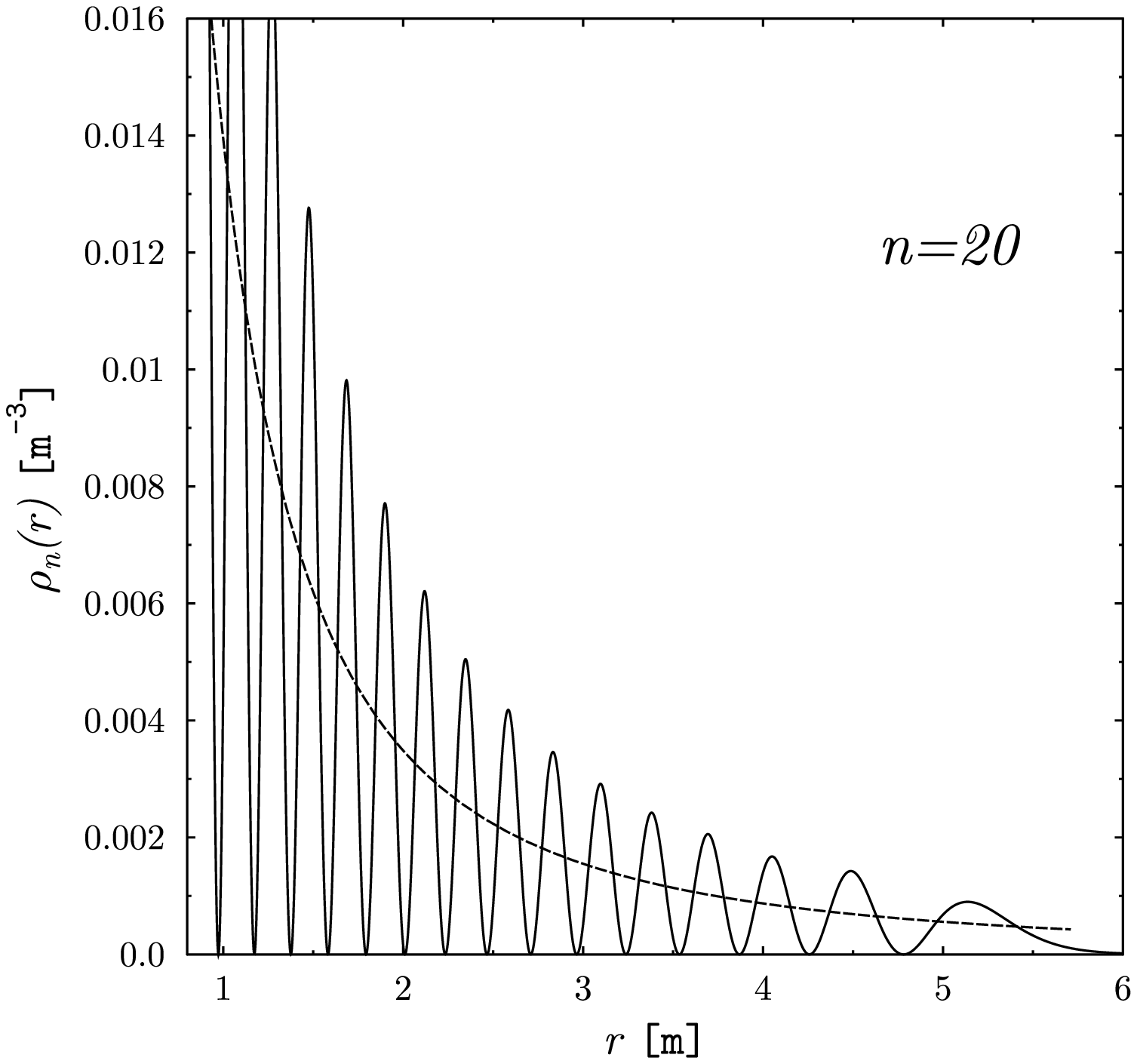}&
\includegraphics[width =.5\textwidth]{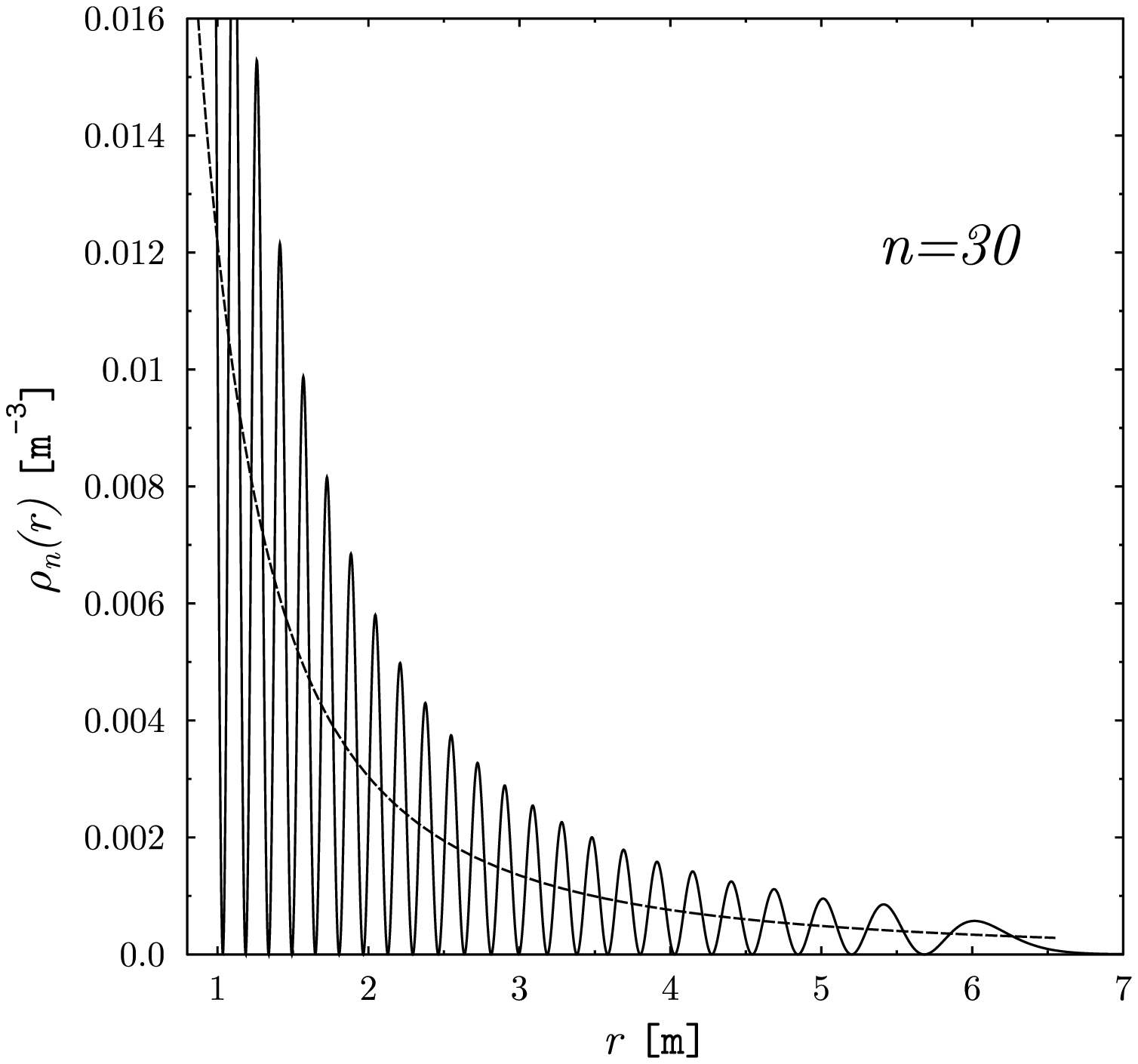}
\end{tabular}
\caption{The plot of quantum probability density
$\rho_n(r)=|\psi_n({\bm x})|^2$ (solid line), where $\psi_n({\bm x})$
is the wave function (3.12) and the classical probability density
$\rho_{\rm cl}(r)$ (dashed line) given by (3.13), where $r_{\rm
max}=r_{\rm max}(E_n)=\frac{\sqrt{2E_n}}{\kappa}$.  We set
$c=1$ m${\rm s}^{-1}$, $\kappa^2=1$ J${\rm m}^{-2}$, and $\hbar=1$ Js.}
\end{figure*}

We now discuss the expectation values of both the kinetic and potential
energies.  Using the identity \cite{23}
\begin{equation}
\frac{1}{[{\rm Ai}'(a_n)]^2}\int_0^\infty x{\rm Ai}^2(x+a_n)dx=
-\frac{2}{3}a_n,
\end{equation}
(3.10) and (3.9) we get
\begin{equation}
\langle \psi_n|c\hat p\psi_n\rangle = c\int d^3{\bm k}|{\bm k}||\psi_n(\bm k)|^2
=\frac{2}{3}E_n,
\end{equation}
where $\hat p=\sqrt{{\hat{\bm p}}^2}$.  Hence, taking into account
the form of the Hamiltonian in Eq.\ (3.1) we find
\begin{equation}
\left\langle \psi_n\Bigg\vert\frac{\kappa^2}{2}{\hat r}^2\psi_n\right\rangle 
= \frac{1}{3}E_n,
\end{equation}
where $\hat r=\sqrt{{\hat {\bm x}}^2}$.  We conclude that the virial
theorem takes the nonstandard form in the case of the massless relativistic
harmonic oscillator.  More precisely, the roles of the kinetic
energy and potential energies are exchanged.  Interestingly, we have
the same formulas on average kinetic and potential energies in the
classical case.  Indeed, from Eq.\ (3.12) it follows easily that
\begin{equation}
\left\langle\frac{\kappa^2}{2}r^2\right\rangle_{\rm cl}=\frac{\kappa^2}{2}
\int_0^{\infty}r^2\rho_{\rm cl}(r)d\mu(r)=\frac{1}{3}E.
\end{equation}
Therefore, by virtue of the first equation of Eq.\ (2.3) we have
\begin{equation}
\langle cp\rangle_{\rm cl} = \frac{2}{3}E.
\end{equation}
We finally write down the following approximate relation obtained
numerically:
\begin{equation}
\langle \psi_n|\hat r\psi_n\rangle\approx \frac{r_{\rm max}(E_n)}
{2}=\frac{\sqrt{2E_n}}{2\kappa},
\end{equation}
where the formula is exact in the limit $n\to\infty$.  The
approximation in (3.19) is very good.  The maximal relative error 
$|(\langle \psi_n|\hat r\psi_n\rangle-r_{\rm max}(E_n)/2)
/\langle \psi_n|\hat r\psi_n\rangle|$ arising in the case with $n=1$
is about 5\% and is lesser than 1\% for $n=2$.  The fact that the
limit $n\to\infty$ when we have the exact equality in (3.19), is the
classical limit is confirmed by the classical formula
\begin{equation}
\langle r\rangle_{\rm cl} = \int_0^{\infty}r\rho_{\rm cl}(r)d\mu(r)=
\frac{r_{\rm max}}{2}=\frac{\sqrt{2E}}{2\kappa},
\end{equation}
following directly from Eq.\ (3.12).
\section{Conclusion}
In this work we study the relativistic massless harmonic
oscillator in both classical and quantum cases.  It seems that the
obtained results concerning the classical oscillator are of
importance not only from the physical point of view.  Indeed, Eq.\ (2.2)
is one of the simplest examples of a nonlinear Hamiltonian system
with constant length of velocity.  As far as we are aware such an
interesting class of nonlinear dynamical systems was not discussed
in the literature.  Referring to the observations of this work
related to the quantum mechanics of the relativistic massless
harmonic oscillator we wish to point out that Eq.\ (3.11) is, to
our best knowledge, the first example of the nontrivial exact solution 
to the Salpeter equation.  We also stress the good behavior of
the corresponding probability density and expectation values of
observables which confirms the correctness of the quantization based 
on the massless spinless Salpeter equation.  Furthermore, we
obtain the exact formula (3.9) on the spectrum of the
Hamiltonian.  It should be noted that for the Salpeter
equation only energy bounds were analyzed in the literature so far
(for the massive relativistic harmonic oscillator see Ref.\ \cite{10}).
Finally, we have obtained the interesting form of the virial theorem for the 
massless relativistic harmonic oscillator with the exchanged roles of 
the kinetic and potential energies.

\end{document}